\documentstyle[epsfig]{aipproc}

\newcommand{\cL}{{\cal L}}
\newcommand{\tr}{{\rm Tr}}
\newcommand{\del}{\partial}
\newcommand{\cD}{{\cal D}}
\newcommand{\cO}{{\cal O}}
\def \lta {\mathrel{\vcenter
     {\hbox{$<$}\nointerlineskip\hbox{$\sim$}}}}

\newcommand{\prd}{{\em Phys.\ Rev.\ }  {\bf D}}

\newcommand{\prl}{{\em Phys.\ Rev.\ Lett.\ }}
\newcommand{\np}{{\em Nucl.\ Phys.\ }{\bf B}}
\newcommand{\pl}{{\em Phys.\ Lett.\ }{\bf B}}

\newcommand{\zp}{Z. Phys.\ {\bf C}}

\begin{document}
\begin{flushright}
MADPH-98-1039\\
%May 22, 1997\\
\end{flushright}
\title{Constraints on Strong Dynamics from Rare B and K Decays
\thanks{Talk presented at the Workshop on Physics at the First Muon Collider
and the Front End of the Muon Collider; November 6-9 1997, Fermilab, 
Batavia, Illinois.}}
\author{Gustavo Burdman}
\address{Department of Physics, University of Wisconsin, Madison 
WI 53706.}

%\lefthead{LEFT head}
%\righthead{RIGHT head}
\maketitle

\begin{abstract}
We discuss the constraints from rare B and K decays on the Electroweak
Symmetry Breaking (EWSB) sector, as well as on theories of fermion masses.
We focus on  models involving new strong dynamics
and show  that  
transitions involving Flavor Changing Neutral Currents (FCNC) play
an important role in disentangling the physics in these scenarios. 
In a model-independent approach to the EWSB sector, the information 
from rare decays is complementary to precision electroweak observables 
in bounding the contributions to the effective lagrangian. 
We compare the pattern of deviations from the Standard Model (SM) 
that results from these sources, with the deviations associated with the 
mechanism for generating fermion masses. 
\end{abstract}

\section*{Introduction}
Two of the most intriguing questions in particle physics are the 
the EWSB mechanism and the origin of fermion masses. 
Although the SM remains a successful theory when compared with all the 
available data, it lacks predictability in the Higgs sector, which determines
the masses of gauge bosons, as well as of fermions through {\em ad hoc} 
Yukawa
couplings. This suggests the possibility that new physics beyond the SM
might be associated with either of these questions. 
In general, the energy scales and dynamics behind the EWSB sector and 
the fermion masses may be unrelated. 
In order to avoid fine-tuning, the scale associated with EWSB cannot be 
much higher than a few~TeV, whereas the scales where light fermion masses are
generated could be much higher.  
If the mechanism responsible for 
the breaking of the electroweak symmetry involves 
some new strong dynamics, deviations from the SM might be 
observable in low energy signals even at energies much smaller than the 
scale of new physics $\Lambda$. 
Reaching this new frontier by direct observation of new physical states
or even of tree-level effects in the couplings of SM particles, may require
not only very large energies but also some previous knowledge of what (and
what not) to expect.
Thus, low energy measurements might be of paramount importance in 
planning experiments and search strategies at high energy machines. 
Among these low energy signals are electroweak measurements such as those 
at LEP and the Tevatron. On the other hand, processes involving 
Flavor Changing Neutral Currents (FCNC) can play a complementary 
role, since the fact that these processes are largely suppressed or forbidden
in the SM may compensate the suppression by  
factors of $m/\Lambda$ (with $m$ the low energy 
scale, e.g. $m_K$, $m_B$, etc.). Here we address the potential of 
rare $B$ and $K$ decays as a complement to other low energy measurements
in constraining models where strong dynamics is associated to either the 
EWSB sector and/or the origin of fermion masses. 
In the absence of a completely satisfactory theory of dynamical symmetry
breaking and fermion masses,  
it is convenient to carry out a model-independent
analysis that makes maximum use of the known properties of the 
electroweak interactions. 
This is the case with the EWSB sector, where an effective
lagrangian approach allows us to parameterize the effects of the 
new strong dynamics in very much the same way chiral perturbation 
theory parameterizes low energy QCD. 
On the other hand, the effects from fermion mass generation 
can also be addressed by a general operator
analysis. However, in addition, most theories predict the existence of 
relatively
light states (scalars, pseudo-Goldstone bosons, etc.) which generally
couple to mass in one way or another. 
To exemplify the effects of such states (which cannot be integrated out)
we work with a particular set of models known as Topcolor-assisted 
Technicolor
(TaTC). This provides a current example of how strong dynamics model 
building deals with the large top-quark mass and illustrates the distinct
low energy phenomenology emerging from non-standard EWSB scenarios.

\section*{Low Energy Effects of Electroweak Symmetry Breaking } 

In the absence of a light Higgs boson the symmetry breaking sector 
is represented by a non-renormalizable effective lagrangian
corresponding to the non-linear realization of the $\sigma$ model. 
The essential feature is the spontaneous breaking of the global symmetry
$SU(2)_L\times SU(2)_R\to SU(2)_V$. To leading order the interactions
involving the Goldstone bosons associated with this mechanism and 
the gauge fields are described by the effective lagrangian~\cite{longhitano}
\begin{equation}
\cL_{LO}=-\frac{1}{4}B_{\mu\nu}B^{\mu\nu}
-\frac{1}{2}\tr\left[W_{\mu\nu}W^{\mu\nu}\right]
+\frac{v^2}{4}\tr\left[D_\mu U^\dagger D^\mu U\right] ,
\label{lo}
\end{equation}
where $B_{\mu\nu}$ and  
$W_{\mu\nu}=\del_\mu W_\nu-\del_\nu W_\mu + ig\left[W_\mu,W_\nu\right]$
are the  the $U(1)_Y$ and $SU(2)_L$ field strengths respectively, 
the electroweak scale is $v\simeq 246$~GeV and 
the Goldstone bosons enter through the matrices
$U(x)=e^{i\pi(x)^a\tau_a/v}$. The covariant derivative acting on $U(x)$
is given by $D_\mu U(x)=\del_\mu U(x)+ igW_\mu (x) U(x) - 
\frac{i}{2}g'B_\mu (x) U(x)\tau_3$. To this order there are no free
parameters once the gauge boson masses are fixed. The 
dependence on the dynamics underlying the strong symmetry breaking
sector appears at next to leading order. 
To this order, a complete set of operators includes one operator of 
dimension two and nineteen operators of dimension four 
\cite{longhitano,appelquist}. 
The effective lagrangian to next to leading order 
in the basis 
of Ref.~\cite{longhitano} is given by 
\begin{equation}
\cL_{\rm eff.}=\cL_{LO} + 
\sum_{i=0}^{19}\alpha_i\cO_i ~~,
\label{lnlo}
\end{equation}
where $\cO_0$ is a
dimension two custodial-symmetry violating term absent in the heavy
Higgs limit of the SM. If we restrict ourselves to CP invariant
structures, there remain fifteen operators of dimension four. 
The coefficients of some of these operators are constrained by low
energy observables. For instance precision electroweak observables
constrain the coefficient of $\cO_0$, which gives a contribution
to the electroweak parameter $T$.
The $3~\sigma$ limit requires 
\begin{equation}
\alpha_0<6\times 10^{-3}. 
\label{alpha0}
\end{equation}
 
The combinations $(\alpha_1+\alpha_8)$ and $(\alpha_1
+\alpha_{13})$ contribute to the electroweak parameters $S$ and $U$.
For instance, the constraint on $S$ translates into
\begin{equation}
|\alpha_1+\alpha_{13}|<1.5\times 10^{-2}.
\label{alpha113}
\end{equation}

In addition, the coefficients $\alpha_2$, $\alpha_3$, $\alpha_9$ and 
$\alpha_{14}$
modify the triple gauge-boson couplings (TGC)
and will be probed at LEPII and the Tevatron at the
few percent level \cite{tgcdef}.

The remaining operators 
contribute to 
oblique corrections only to one loop and, in some cases, only starting at two
loops. 
To the last group belong $\cO_{11}$ and $\cO_{12}$ given that 
their contributions to the gauge boson two-point functions only affect
the longitudinal piece of the propagators. 
Of particular interest is 
the operator $\cO_{11}$ defined by \cite{longhitano}
\begin{equation}
\cO_{11}=\tr\left[\left(\cD_\mu V^\mu\right)^2\right], 
\label{defo11}
\end{equation}
with $V_\mu=(D_\mu U)U^\dagger$ and the covariant derivative acting on
$V_\mu$ defined by $\cD_\mu V_\nu=\del_\mu V_\nu +
ig\left[W_\mu,V_\nu\right]$. The equations of motion for the
$W_{\mu\nu}$ field strength imply \cite{feruglio}
\begin{equation}
\cD_\mu V^\mu = \frac{2i}{v^2}~\cD_\mu J_{w}^{\mu}~, 
\label{equmot}
\end{equation}
where the $SU(2)_L$ current is $J_{w}^\mu=\sum_{\psi}\left(\bar{\psi}_L
\gamma^\mu \frac{\tau^a}{2}\psi_L\right)\tau^a$, $\psi_L$ denote the 
left-handed fermion doublets. 
The dominant effect 
appears in the quark sector due to the presence of terms proportional to 
$m_t$. After the quark fields are rotated to the mass eigenstate basis, the
operator $\cO_{11}$ can be written as \cite{pich}
\begin{equation}
\cO_{11}=\frac{m_t^2}{v^4}\left\{\left(\bar{t}\gamma_5 t\right)^2
-8\sum_{i,j}V_{ti}^*V_{tj}(\bar{q_i}_L t_R)(\bar{t}_R q_{j L})\right\}
+ \dots 
\label{ffo11}
\end{equation}
where $i,j=d,s,b$, the $V_{ti}$ are Cabibbo-Kobayashi-Maskawa
(CKM) matrix elements and 
the dots stand for terms suppressed by small fermion masses.

From the above discussion we see that the leading effects of the EWSB 
sector in FCNC processes are coming from the insertion of 
anomalous TGC vertices and four-fermion operators like (\ref{ffo11}). 
In the rest of this section, we review the status and future impact
of these constraints on the symmetry breaking sector.

\subsection*{Four-fermion Operators}  
The effects of the four-fermion operators in (\ref{ffo11}) 
in rare B and K decays were considered in Ref.~\cite{bews}. 
The loop insertion will result in contributions to several FCNC processes, 
that are controlled by both the coefficient $\alpha_{11}$ of the effective
lagrangian (\ref{lnlo}) as well as by the high energy scale $\Lambda$. 
To one loop, only one parameter is needed, namely
\begin{equation}
y=\alpha_{11}\log\frac{\Lambda^2}{m_t^2}~.
\label{ydef}
\end{equation}
This parameter also governs the contributions of (\ref{ffo11}) to  
other neutral processes, both flavor changing and flavor conserving. 
For instance, the $(\bar b_Lt_R)(\bar t_Rb_L)$ term in (\ref{ffo11})
gives a contribution to $Z\to b\bar b$, whereas the 
terms like $(\bar b_Lt_R)(\bar t_Rd_L)$ appear in $B^0-\bar{B^0}$
mixing~\cite{pich}.
Thus the measurements of $R_b$ and the rate of $B$ mixing (together
with all other CKM information) can be used to derive a bound on 
$y$. Although the bound carries some uncertainty mainly associated
with CKM quantities like $f_B$ and $V_{ub}$, we will take it to be, 
approximately~\cite{pich,bews}
\begin{equation}
|y|<0.50~.
\label{ybound}
\end{equation}
Next, we use this as the allowed range for $y$ in order to explore the 
possible impact of this physics in rare $B$ and $K$ decays. 
The one-loop insertion of the terms 
\begin{equation}
\cO_{11}=-\frac{8m_t^2}{v^4}\left\{V_{ts}^*V_{tb}\bar{s}_Lt_R\bar{t}_Rb_L
+V_{td}^*V_{tb}\bar{d}_Lt_R\bar{t}_Rb_L+V_{td}^*V_{ts}\bar{d}_Lt_R
\bar{t}_Rs_L
\right\} + \dots
\label{rareo11}
\end{equation}
induces new contributions to various FCNC vertices 
in $B$ decays (the first two terms in (\ref{rareo11})), 
as well as in $K$ decays (third term in (\ref{rareo11})).  

First, let us consider $b\to q\gamma$ processes leading, for instance, to 
the inclusive $B\to X_s\gamma$, since this rate has been recently 
measured~\cite{bsgexp}. The one-loop insertion of the operator $\cO_{11}$
does not give a contribution to these processes given that it does not
mix with the operator $\bar s_L\sigma_{\mu\nu}b_R$ responsible for the 
on-shell photon amplitude. Mixing only occurs at two loops, when QCD 
corrections are taken into account. As a result the effect, 
in all $b\to q\gamma$ transitions is expected to be only a few percent 
of  the SM branching ratios~\cite{bews}.  

\begin{figure}[t!] % fig 1
\centerline{\epsfig{file=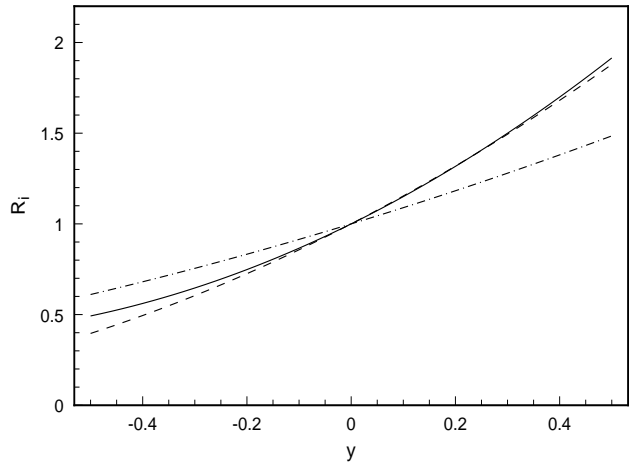,height=2.5in,width=3.5in}}
\vspace{10pt}
\caption{Ratio of the modified branching ratio to the standard model
expectation as a function of $y=\alpha_{11}\log{\frac{\Lambda^2}{m_t^2}}$.
The solid line corresponds to the ratio $R_\ell$ for 
$B\to X_{(s,d)}\ell^+\ell^-$ inclusive decays, the dashed line
to $R_\nu$ for $B\to X_{(s,d)}\nu\bar\nu$ and the dot-dashed line to
$R_g$ for $b\to s\bar s s$ decays. From Ref.~[6].}
\label{fig1}
\end{figure}

On the other hand, the off-shell
amplitudes for photons, $Z$'s and gluons are non-zero at one loop. 
They generate contributions to processes such as $b\to q\ell^+\ell^-$, 
$b\to q\nu\bar\nu$, $b\to q\bar q' q'$; as well as to similar 
rare kaon decays like $s\to d\nu\bar\nu$, etc. In order to asses the 
potential effects we define 
\begin{equation}
R_\ell\equiv\frac{Br(B\to X_{(s,d)}\ell^+\ell^-)}
{~~\;\;Br(B\to X_{(s,d)}\ell^+\ell^-)_{\rm SM}} ~,
\label{defrat}
\end{equation}
which is plotted in Fig.~1 as a function of the parameter $y$ defined in 
(\ref{ydef}), for the allowed range of $y$ (\ref{ybound}). 
Analogously, we can define
the ratio $R_\nu$, which tracks the effects in $B\to X_{s,d}\nu\bar\nu$
decays; whereas the contribution to gluon penguin processes such as 
$b\to s\bar ss$ is represented by the ratio $R_g$. As it is clear from 
Fig.~\ref{fig1}, the effects of the operator $\cO_{11}$ are very similar 
in all three types of $B$ decays.

We see that, even with the $R_b$ and $B^0-\bar B^0$ mixing constraints,  
large deviations from the SM predictions for these modes are possible. 
The current experimental bounds on these processes are still not binding
on $y$. However, sensitivity to SM branching ratios will be reached in the 
next round of experiments at the various $B$ factories at Cornell, KEK, SLAC
and Fermilab. The distinct feature of this effect is that no significant
deviation is expected in $b\to s\gamma$, even when large deviations
are observed in all the other modes. 

The effects are very similar in rare $K$ decays such as 
$K^+\to\pi^+\nu\bar\nu$ and $K_L\to\pi^0\nu\bar\nu$, etc. In Fig.~2
we plot $R_K$, a
quantity analogous to $R_\ell$ in (\ref{defrat}). Again, large effects of 
up to factors of $2$ deviations, are allowed. The recently
reported~\cite{kpinu} observation of one event in $K^+\to\pi^+\nu\bar\nu$
roughly translates into $R_K\simeq (0.50-5.0)$, which is still not
constraining. 

Although in this model-independent approach we cannot, as a matter of
principle, calculate the size of the coefficients $\alpha_i$, we can use 
general arguments to estimate their approximate value.
Using naive dimensional analysis~\cite{nda} we have
\begin{equation}
\alpha_{11}\simeq {\cal O}(1)\times\frac{v^2}{\Lambda^2}~,
\label{da}
\end{equation}
with the scale of new physics obeying $\Lambda\lta 4\pi v$. 
For instance, taking $\Lambda=4\pi v$, 
one would obtain
$y\simeq{\cal O}(1)\times0.04$. On the other hand, if $\Lambda=2\pi v$, 
one has $y\simeq{\cal O}(1)\times0.12$. In any case, these are meant to be
order of magnitude estimates.
Therefore, the experimental relevance
of the effect strongly depends on details of the dynamics we are not 
able to compute in a model-independent fashion. 

Finally, we should note that rare $B$ and $K$ decays are the most sensitive
signals for this effect. This is due to the fact that four-lepton operators
are suppressed by the lepton masses, and that $\cO_{11}$ does not mix
quarks and leptons.

\begin{figure}[t!] % fig 2
\centerline{\epsfig{file=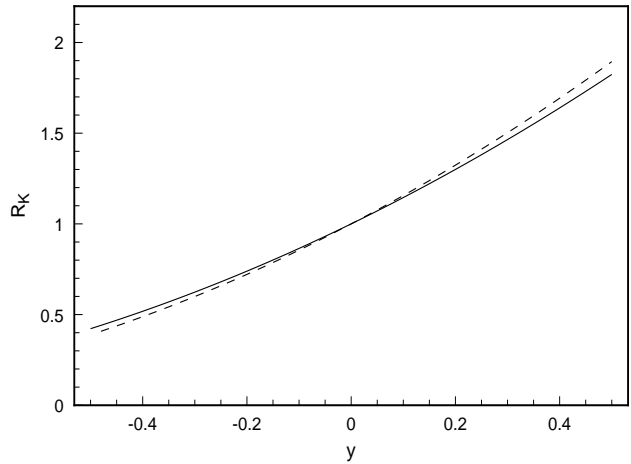,height=2.5in,width=3.5in}}
\vspace{10pt}
\caption{Ratio of the modified branching ratio to the standard model
expectation for $K^+\to\pi^+\nu\bar\nu$ (solid line) and 
$K_L\to\pi^0\nu\bar\nu$ (dashed  line). From Ref.~[6].}
\label{fig2}
\end{figure}

\newpage
\subsection*{Triple Gauge-boson Couplings}
Imposing $C$ and $P$ conservation, the most general form of the 
$WWN$ ($N=\gamma, Z$)
couplings can be written as~\cite{tgcdef}
\begin{eqnarray}
\cL_{WWN}&=&g_{WWN}\left\{i\kappa_N W_{\mu}^{\dagger}W_\nu N^{\mu\nu}
+ig_1^N \left(W_{\mu\nu}^{\dagger}W^\mu N^\nu -
W_{\mu\nu}W^{\dagger\mu} N^\nu\right) \right. \\
& & \left. +i\frac{\lambda_N}{M_W^2}W^\dagger_{\mu\nu}W^\nu_{~\lambda} 
N^{\nu\lambda}
\right\}~, 
\label{wwnc}
\end{eqnarray}
with the conventional choices being $g_{WW\gamma}=-e$ and 
$g_{WWZ}=-g\cos\theta$~\cite{tb}. 
In principle, there are six free parameters. 
Making contact with the electroweak lagrangian (\ref{lnlo}), 
these parameters can be expressed in terms of the next-to-leading
order coefficients~\cite{feruglio}
$\alpha_1,\alpha_2,\alpha_3,\alpha_8,\alpha_9,
\alpha_{13}$ and $\alpha_{14}$. 
Conservation of the electromagnetic charge  implies $g_1^\gamma=1$. 
Furthermore, to this order in the energy expansion (\ref{lnlo})
$\lambda_N=0$.
Then we are left with $\kappa_\gamma$, $\kappa_Z$ and $g_1^Z$. 
Finally, when considering rare $B$ and $K$ decays, we can neglect the 
contribution of $\kappa_Z$ since it will be suppressed by powers of the 
small external momenta over $m_Z$. 
Thus, in this simplistic approach, there are only two parameters
relevant at very low energies. The SM predicts $\kappa_\gamma=g_1^Z=1$.  
The effects of anomalous TGC 
have been previously studied in the literature~\cite{tgcinb}.
However, this hierarchical approach to the couplings has not been the one 
used in the various analyses and a more comprehensive study is needed. 
The experiments at LEP~II and the next Tevatron run are going
to be sensitive to deviations from the SM prediction at the 
$(5-10)\%$ level~\cite{tb}. 
Effects of this size might be also observed in rare
$B$ and $K$ decays. For instance, $\delta g_1^Z=g_1^Z - 1=0.10$ 
can produce enhancements in the branching ratios of 
$b\to s\ell^+\ell^-$ decay modes of up to $(60-70)\%$~\cite{tgcinb}. 
In the near future, $B$ factory experiments will have sensitivity 
to these processes at the SM level, turning these low energy measurements 
into an excellent complement of direct probes of the TGC. 

\section*{Fermion Masses and Electroweak Dynamics}
Up to now, we have only considered the effects of the dynamics associated
with the EWSB. These are encoded in the effective lagrangian (\ref{lnlo}), 
which only involves the Goldstone boson and gauge fields. 
Additionally, it is possible that the new strong dynamics may also affect
some or all fermions. 
We first comment on the effective lagrangian approach
for non-SM couplings of fermions to gauge bosons, 
and then examine the effects of a prototypical class of 
theories (Topcolor) where the dynamical generation of fermion masses
imply the existence of relatively light new states.
  
\subsection*{Anomalous Couplings of Fermions to Gauge Bosons}
The effects of new dynamics 
on the couplings of fermions with the SM gauge bosons can be, in principle, 
also studied in an effective lagrangian approach.  For
instance,  
if in analogy with the situation in QCD, fermion masses are
dynamically generated in association with EWSB, 
residual interactions of fermions with Goldstone bosons 
could be important~\cite{peccei} if the $m_f\simeq f_{\pi}\simeq v$. 
Thus residual, non-universal interactions of the third generation
quarks with gauge bosons could carry interesting  information about
both the origin of the top quark mass and EWSB. 
In a very general parameterization, the anomalous couplings of third 
generation
quarks can be written as
\begin{eqnarray}
\Delta \cL&=&-\frac{g}{\sqrt{2}}\left\{ C_L\;(\bar t_L\gamma_\mu b_L)
+C_R\;(\bar t_R\gamma_\mu b_R)\right\}W^{+\mu} \nonumber\\
& &-\frac{g}{2\;c\theta_W}\left\{ N_L^t \;(\bar t_L\gamma_\mu t_L) 
+ N_R^t \;(\bar t_R\gamma_\mu t_R)\right. \nonumber\\
& &\left.+ N_L^b \;(\bar b_L\gamma_\mu b_L) 
+ N_R^b \;(\bar b_R\gamma_\mu b_R)
\right\}Z^\mu~,
\label{lfer}
\end{eqnarray}
where the parameters $C_{L,R}$, $N_{L,R}^{t,b}$ contain the residual, 
non-universal effects associated with the new dynamics, perhaps responsible
for the large top quark mass.
Then, if we assume that the new couplings are CP conserving, there are 
six new parameters. They are constrained at low energies by 
a variety of experimental information, mostly from electroweak precision
measurements and the rate of $b\to s\gamma$. 
Several simplifications are usually made in order to reduce the number of 
free parameters. For instance, 
in most of the literature, it is assumed that $N_{L,R}^b=0$~\cite{cpyuan}. 
A stringent bound on the right-handed charged coupling is obtained from
$b\to s\gamma$~\cite{fuji}: $-0.05<C_R<0.01$. 
The bounds obtained on a particular coupling 
from electroweak observables such as $S$, $T$, $U$
and $R_b$ generally strongly depend on assumptions about the other 
couplings. 
For example, if $C_L=0$, then the combination $(N_L^t-N_R^t)$ is strongly
constrained since it contributes to $T$. On the other hand, if 
$C_L=N_L^t$, then  $N_R^t<0.02$~\cite{peccei,bdjt} since it is 
the only (linear) contribution to $T$. 
Thus, although in general most parameters are confined to a few percent, 
some of them are allowed to be as large as $0.30$ under certain conditions. 
This ``model-dependent'' situation requires more experimental information. 
A global analysis of the effects of the couplings of eqn.~(\ref{lfer})
in rare $B$ and $K$ processes such as $b\to s\ell^+\ell^-$, 
$s\to d\nu\bar\nu$, etc. may help disentangle the various possible 
effects and perhaps will give constraints that may be of importance in
interpreting data from higher energy experiments~\cite{meprep}.

\subsection*{The effects of light states: the example of Topcolor}
The description of the residual effects of strong dynamics at low energies
on fermion couplings by using (\ref{lfer}) corresponds to cases where the 
states associated with the new physics are heavy compared to the weak scale.
Thus, integrating out the heavy states, leaves us with effective couplings
which might be generated at tree level or through loops in the full theory. 
However, most theories in which electroweak symmetry and/or fermion
masses have a dynamical origin also contain states with masses comparable
to the weak scale. Such is the case, for instance, in Technicolor models
where the breaking of large chiral symmetries imply the presence of 
pseudo-Goldstone bosons with masses of at most a few hundred GeV. 
It is also the case in Topcolor-assisted Technicolor (TaTC) 
models~\cite{tc2}, 
where a top-condensation mechanism generated by the Topcolor interactions
is responsible for the large dynamical top quark mass, whereas Technicolor
breaks the electroweak symmetry giving (most of) the $W$ and $Z$ masses.
The TaTC scenario is designed to relief the problems of Extended 
Technicolor (ETC) in generating a heavy top~\cite{el}. 
Although the new gauge bosons associated with the TaTC gauge group are 
heavier than $1~$TeV, the presence of several scalar and pseudo-scalar
states with masses in the few-hundred~GeV range, forces us to take these 
into account
directly in our calculations. 
From the point of view of their impact 
in low energy observables, the most important of these states are the 
top-pions $\vec\pi_t$, the triplet of Goldstone bosons associated  with the 
breaking of the top chiral symmetry. Since top condensation does not fully
break the electroweak symmetry ($f_{\pi_t}\simeq (60-70)~{\rm GeV}<v$), 
after mixing with the techni-pions, there will be a triplet of physical
top-pions in the spectrum, with a coupling to third generation quarks 
given by
\begin{equation}
i\frac{m_t}{\sqrt{2}f_{\pi_t}}\left\{ \bar t\gamma_5 t \pi^0 
+\bar t_Rb_L \pi^+ +\bar b_L t_R \pi^- \right\}~.
\label{tpcop}
\end{equation}
They acquire masses of a few hundred GeV due to explicit ETC 
quark mass terms. 
Additionally, in most models there are scalar and pseudo-scalar bound states 
due to the strong (although sub-critical) effective coupling of 
right-handed $b$-quarks. The closer the effective couplings are 
from criticality,
the lighter these bound states tend to be. 
The spectrum and properties of these states,  
unlike those of top-pions, are not determined by model-independent features
of the symmetry breaking pattern but depend 
on details of the model. 
Finally, in all TaTC models there will be pseudo-Goldstone bosons 
from the breaking of techni-fermion chiral symmetries. However, 
their couplings to third generation quarks are reduced with respect
to (\ref{tpcop}) by $m_{ETC}/m_t$, where $m_{ETC}$ is a small ETC
mass of the order of $1~$GeV\footnote{Multi-scale Technicolor models such 
as the one in Ref.~\cite{multi}, in the absence of Topcolor, 
have un-suppressed top couplings to pseudo-scalars.
This could lead to effects similar to those of top-pions.}.
The presence of the relatively light top-pions, as well as the additional
bound states, imposes severe constraints on Topcolor models due to their 
potential loop effects in low energy observables, most notably $R_b$ and 
rare $B$ and $K$ decays. 

\vspace{0.3cm}
\paragraph*{Top-pion Effects in $R_b$}:~ The one-loop contributions 
of top-pions to the $Z\to\bar b b$ process were studied in Ref.~\cite{gbdk}.
There it was shown that they shift $R_b$ negatively by an amount controlled
by $m_{\pi_t}$ and $f_{\pi_t}$. For instance, for $f_{\pi_t}\simeq 60~$GeV
the correction is about $-1\%$ for $m_{\pi_t}=800~$GeV, and top-pions 
with masses in the expected $(100-300)~$GeV range give unacceptably large
deviations. This value of the top-pion decay constant is obtained by 
using the Pagels-Stokar formula, which gives $f_{\pi_t}$ a logarithmic
dependence on the Topcolor energy scale, chosen here to be a few TeV. 
Potentially cancelling contributions by other states, 
such as the scalar and pseudo-scalar bound states, Topcolor vector and 
axial-vector mesons, etc., are either of the wrong sign or not large enough. 
Possible ways out of this constraint are: larger top-pion masses 
or larger values of $f_{\pi_t}$. The larger
$f_{\pi_t}$ is, the smaller the coupling, and the top-pions are more
Goldstone-boson-like. For $f_{\pi_t}\simeq 120~$GeV, for instance, the 
shift of $R_b$ is well within the experimentally allowed region even for 
$m_{\pi_t}\simeq (200-300)~$GeV. However, in order to obtain such an 
enhancement in the decay constant we must either assume large corrections
to the Pagels-Stokar expression or introduce new and exotic fermion states.

\vspace{0.2cm}
\paragraph*{Rare $B$ and $K$ Decays}:~
The top-pions and other scalar states, give one-loop contributions
to FCNC processes. These depend not only on $f_{\pi_t}$ and $m_{\pi_t}$
but typically also on one or more elements of the quark rotation 
matrices necessary to diagonalize the quark Yukawa couplings. 
The contributions of top-pions, as well as ``b-pions'' (scalar and 
pseudo-scalar bound states in models where $b_R$ couples to the Topcolor
interaction) to $b\to s\gamma$ depend on $D_{L,R}^{bs}$, the $b\to s$ 
element in the left or right down rotation matrix. Furthermore, the two 
contributions tend to cancel. Thus, the freedom in this model-dependent 
aspects of the prediction makes it possible to have quite low 
masses  and still satisfy the bound from the experimental measurement
of $B\to X_s\gamma$~\cite{bbhk}.
The situation changes drastically in $b\to s\ell^+\ell^-$ processes, 
where the cancellations are much less efficient.
Although experiments have not yet reached sensitivity to SM branching 
ratios~\cite{bsllexp}, it will be soon achieved at both hadron and lepton
$B$ factories. As an example, we plot in Fig.~3 the 
Br$(b\to s\ell^+\ell^-)$ 
as a function of the top-pion mass with no other contributions, for 
$f_{\pi_t}=70~$GeV.
\begin{figure}[t!] % fig 3
\centerline{\epsfig{file=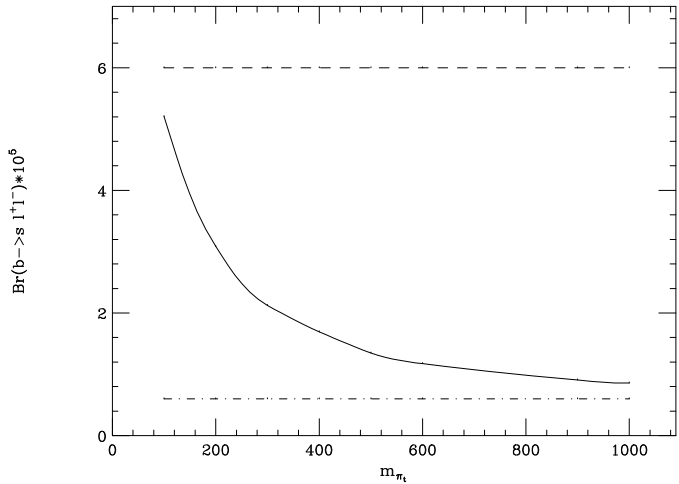,height=3.5in,width=2.5in,angle=90}}
\vspace{10pt}
\caption{Br($b\to s\ell^+\ell^-$) vs. $m_{\pi_t}$, for $f_{\pi_t}=70~$GeV.
The dashed horizontal line is the current experimental limit for the 
inclusive rate~[24], 
whereas the dot-dash line is the SM expectation.}
\label{fig3}
\end{figure}
The 
$b\to s\gamma$ constraint is in this case
$(200~{\rm GeV}<m_{\pi_t}<800~{\rm GeV})$. However, one can see  that, 
even for
heavier top-pions the effect can still be a $(30-60)\%$ enhancement over the
SM prediction of $6\times 10^{-6}$. 
On the other hand, in the presence of a $400~$GeV charged b-pion the curve
changes little, but the $b\to s\gamma$ bound is now 
$m_{\pi_t}<600~{\rm GeV}$.
Finally, to compare the potential of these FCNC transitions with the $R_b$
constraints, let us say that if we take $f_{\pi_t}\simeq 120~$GeV (which
avoids conflict with $R_b$ measurements), then the effect of a $400~$GeV
top-pion in $b\to s\ell^+\ell^-$ is still an enhancement of more than 
$50\%$ with respect to SM expectations.
Thus, the observation of these modes  will further constrain Topcolor 
models beyond the $R_b$ bounds. 
We expect similar effects due to top-pions and/or b-pions to be present
in kaon processes such as $K^+\to\pi^+\nu\bar\nu$. 

\section*{Conclusions}
We have seen that a complete, model-independent analysis of the effects
of strong dynamics in rare $B$ and $K$ decays could shed light on the nature
of the EWSB mechanism and the origin of fermion masses. 
The signals are also likely to be important in models where relatively light 
scalars couple strongly to mass, like in the case of TaTC. 
In most cases, the next round of experiments will have sensitivity to SM
branching ratios. This will be the case, for instance, for the 
Tevatron experiments,  as well the KEK and SLAC $B$ factories in the 
$B\to X_{(s,d)}\ell^+\ell^-$ modes. It will also be the situation in the 
next generation of kaon experiments for $K^+\to\pi^+\nu\bar\nu$ and 
$K_L\to\pi^0\nu\bar\nu$. 
The amount and variety of experimental information from these processes is 
such that suggests a parallel to the role of electroweak measurements 
at the $Z$ pole as not only a constraint on new physics sources but also 
as guidance in the searches to be carried out at high energy machines such 
as the Tevatron in Run~II, the LHC and eventually the NLC and/or the 
muon collider. It is possible to imagine a scenario where deviations
from the SM in $B$ and/or $K$ decays point to a particular source, 
e.g. corrections to Goldstone boson propagators given by $\cO_{11}$, 
anomalous TGC or anomalous couplings of third generation quarks to gauge
bosons as  in  (\ref{lfer}).  The nature of the deviation might dictate
the road to follow at high energies. As an example, if the source 
of an effect is in 
one the top quark couplings $N_{L,R}^t$, there would be a strong case for 
a lepton collider running at $t\bar t$ threshold. Other scenarios may not
be so clear, and may require a comprehensive and careful analysis of all the 
data to come (including issues like hadronic uncertainties in $B$ decays). 
This, however, constitutes a very well defined research program.

\end{document}